\documentclass[preprint2]{aastex}

\shorttitle{Gas attractor}
\shortauthors{Juncher, Hansen \& Macci\`o}

\begin{document}


\title{An attractor for the dynamical state of the intracluster medium}



\author{Diana Juncher~\altaffilmark{1},
Steen H. Hansen~\altaffilmark{1}, \&
Andrea V. Macci\`o~\altaffilmark{2}}

\altaffiltext{1}{Dark Cosmology Centre, Niels Bohr Institute, University of Copenhagen,
Juliane Maries Vej 30, 2100 Copenhagen, Denmark}
\altaffiltext{2}{Max Planck Institut f\"{u}r Astronomie,  
K\"{o}nigstuhl 17, 69117 Heidelberg, Germany }

%




\begin{abstract}
{Galaxy clusters provide us with important information about the
cosmology of our universe. Observations of the X-ray radiation or of
the SZ effect allow us to measure the density and temperature of the
hot intergalactic medium between the galaxies in a cluster, which then
allow us to calculate the total mass of the galaxy cluster.  However,
no simple connection between the density and the temperature profiles
has been identified.  Here we use controlled high-resolution
numerical simulations to identify a relation between the density and
temperature of the gas in equilibrated galaxy clusters. We demonstrate
that the temperature-density relation is a real attractor, by
showing that a wide range of equilibrated structures all move towards
the attractor when perturbed and subsequently allowed to relax.  For
structures which have undergone sufficient perturbations for this
connection to hold, one can therefore extract the mass profile
directly from the X-ray intensity profile.}
\end{abstract}





\section{Introduction}
Galaxy clusters are the largest known gravitationally bound structures
in the universe. Consisting almost entirely of dark matter and
intracluster gas, they have proved to be very useful objects of study
in cosmology. Cluster observations have imposed constraints on the
parameters of the $\Lambda$CDM model \citep{2011ARA&A..49..409A, 2009ApJ...692.1060V}, and gravitational
lensing reveals the galaxies of the distant universe behind the
clusters \citep{2004ApJ...607..697K}. Observations of the gas in galaxy clusters
have enabled us to quantify the density profile of dark matter
\citep{2004ApJ...604..116B,2005A&A...435....1P,2011ApJ...736...52H}. 
However, both X-ray and SZ intensity
observations suffer from the complication that whereas it is easy to
observe the intensity, which is a product of density and temperature,
then it is much more complicated to observe the temperature and
density separately~\citep{1986RvMP...58....1S}.  Alternatively, if a simple
relation between the density and temperature would exist, one could
extract the total mass profile directly from the easily observable X-ray
or SZ intensities.

Analytically one might try to search for such a relation by turning to
the equation of hydrostatic equilibrium \citep{1976A&A....49..137C}. 
This equation relates the
total mass of a cluster $M_{tot}$ to the gas density $\rho_g$ and
temperature $T_g$ of the intracluster gas. It is given by
\begin{equation}
  M_{tot}(r) = -r \frac{k_BT_g(r)}{\mu m_u G} \left(\gamma_g(r) + \tau_g(r) \right),
\label{eqn:he}
\end{equation}
where $k_B$ is the Boltzmann constant, $\mu$ is the mean molecular
mass, $m_u$ is the atomic mass unit, $G$ is the gravitational
constant and we define $\gamma_g =
\textrm{d}\ln(\rho_g)/\textrm{d}\ln(r)$ and $\tau_g =
\textrm{d}\ln(T_g)/\textrm{d}\ln(r)$. Unfortunately, 
for a given $M_{tot}$ we can choose essentially any
random $\rho_g(r)$ and still make the equation hold by finding a suitable
$T_g(r)$.

In this Letter we study the evolution of galaxy clusters using
hydrodynamical simulations. We find that a wide range of initially 
equilibrated
structures, 
when dynamically perturbed in order to mimic real perturbations
happening during structure formation,
all move towards
a particular curve in the 2-dimensional phase space spanned by the
logarithmic derivatives of the density and temperature of the gas.

\section{Initial Conditions}
The dimensions of our structures are chosen to resemble those of a
large galaxy cluster such as the Perseus Cluster. Each cluster has a
scale radius of $r_0 = 100$kpc and a total mass of $10^{15}$M$_\odot$
of which 10\% is gas and 90\% is dark matter. They are described by 1
million particles of which 10\% represent the gas and 90\% represent
the dark matter. The mass of the gas and the dark matter particles are
thus the same, namely $10^9$M$_\odot$.

To obtain a wide range of initial conditions, we create 20 different
structures that vary in both gas and dark matter density profiles as
well as dark matter velocity distributions. We use double-power laws
to describe the density profiles of the dark matter and the gas.
\begin{equation}
\rho(r) = \rho_0\frac{1}{\left( \frac{r}{r_0}\right)^\alpha} \frac{1}{\left(1 + \frac{r}{r_0}\right)^{\delta-\alpha}}
\end{equation}
Here $\alpha$ and $\delta$ correspond to the inner and outer logarithmic 
slopes of the density profile respectively. The logarithmic slopes of
the dark matter density profiles are chosen in the range $-2 \leq
\gamma_{dm} \leq 0$ in the inner region and $-6 \leq \gamma_{dm} \leq
-3.5$ in the outer region. For a given dark matter density profile we
use the Eddington method to create isotropic dark matter velocity
distributions with $\beta = 0$, or the Osipkov-Merrit method to create
anisotropic dark matter velocity distributions with $\beta(r) =
1/(1+(r_a/r)^2)$, where $r_a = r_0$ or $r_a = 2r_0$
\citep{2008gady.book.....B}. The logarithmic slopes of the gas density profiles
are chosen in the range $-1.5 \leq \gamma_g \leq 0$ in the inner
region and $-5 \leq \gamma_g \leq -3.5$ in the outer region. For a
specific combination of gas and dark matter density profiles we use
the equation of hydrostatic equilibrium to find the corresponding gas
temperature profiles. Note that $\gamma_g$ and $\tau_g$ are
independent variables. It is therefore, in principle, possible to populate much of 
the space spanned by $\gamma_g$ and $\tau_g$.

\begin{figure}[h!]
  \resizebox{\hsize}{!}{\includegraphics{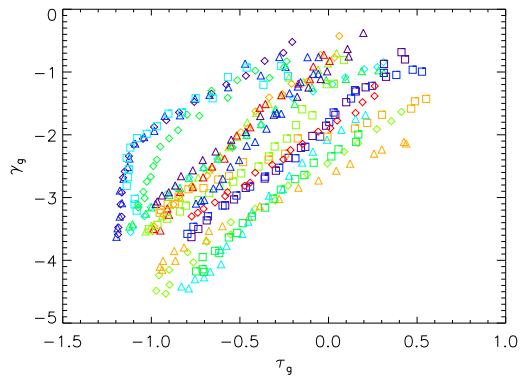}}
  \caption{The initial conditions in the $\gamma_g =
    \textrm{d}\ln(\rho_g)/\textrm{d}\ln(r)$ and $\tau_g =
    \textrm{d}\ln(T_g)/\textrm{d}\ln(r)$ space.  The variations in gas
    and dark matter density profiles as well as dark matter velocity
    distributions allow the structures to span a large portion of the
    $\tau_g$-$\gamma_g$ space. All the structures of 
    this figure are stable on a timescale of 2 dynamical times
    when evolved in a hydrodynamical simulation.}
  \label{fig1}
\end{figure}

All of these initial conditions are presented in Figure \ref{fig1}. To
ensure stability these structures have all been evolved for 2 dynamical times,
where the dynamical time $t_{dyn}$ = 1.47Gyr has been computed at $R=10r_0$.
In all figures we only show the region between 3 times the gravitational softening 
and 10 times the scale radius. The values for $\tau_g$ and $\gamma_g$ are 
the average of particles in logarithmically spaced radial bins. We emphasise that our initial 
conditions both include cool-core clusters with a central $\tau_g > 0$, as well as non-cool 
core clusters with a central $\tau_g < 0$. All these structures are stable on a timescale of 2 dynamical
times.

\section{Numerical Code}
We use the parallel TreeSPH code GADGET-2 to evolve the structures
\citep{2005MNRAS.364.1105S}. GADGET-2 simulates fluid flows by means of
smoothed particle hydrodynamics and computes gravitational forces with
a hierarchical tree algorithm. The time integration is based on a
quasi-symplectic scheme, where long-range and short-range forces can
be integrated with different time steps. 
Radiative cooling is implemented for a primordial mixture of hydrogen and helium 
following \citet{1996ApJS..105...19K}.

The SPH properties are smoothed over the standard GADGET-2 kernel
using $50 \pm 1$ SPH particles, while the gravitational forces are
adjusted by the gravitational spline kernel using a softening length
of 6.0kpc. To protect against too large time steps for particles with
very small accelerations, we require that the maximum time step is two
percent of the dynamical time of the system. We use an artificial
viscosity parameter of $\alpha_{visc} = 0.8$ and a courant factor of
$\alpha_{cour} = 0.1$.

For modelling star formation and the associated heating by supernovae
(SN) we follow the sub-resolution multiphase ISM model developed by
\citet{2003MNRAS.339..289S}. In this model, a thermal instability is assumed
to operate above a critical density threshold $\rho_{th}$, producing a
two-phase medium which consists of cold clouds embedded in a tenuous
gas at pressure equilibrium. Stars are formed from the cold clouds on
a timescale chosen to match observations \citep{1998ApJ...498..541K} and
short-lived stars supply an energy of $10^{51}$ergs to the
surrounding gas as SN.
We adopt the standard parameters for the multiphase model in order to
match the Kennicutt Law as specified in \citet{2003MNRAS.339..289S}. The star
formation timescale is set to $t_*^0 = 2.1$Gyr, the cloud evaporation
parameter to $A_0 = 1000$ and the SN ``temperature'' to $T_{\rm  SN}=10^8{\rm}$K.
More information on the GADGET version adopted in this work 
can be found in \citet{2011MNRAS.415.3750M}.
All simulations are carried out in a non-cosmological Newtonian box.

\section{Perturbations}
The cosmological structures we observe today have a long history of hierarchical
structure formation  involving gravitational collapse and
mergers. Both of these are violent relaxation processes that change
the gravitational potential and therefore also the energy of the
particles according to
\begin{equation}
\frac{\textrm{d}E}{\textrm{dt}} = \frac{\partial \Phi}{\partial t},
\end{equation}
for spherical systems, where $E$ is the energy, $\Phi$ is the
gravitational potential and $t$ is the time 
\citep{1967MNRAS.136..101L, 2008gady.book.....B}

To mimic these processes we will expose all our initially equilibrated
structures to controlled perturbations. We will require that these
perturbations are both continuous and spherical.  One example of such
perturbations is to let the value of the gravitational constant $G$
vary in the N-body code~\citep{sparre12}. The perturbations are
spherical since the structures themselves are spherical. They are also
continuous since they affect the accelerations and not the velocities
of the particles.

Initially, we change the value of $G$ by 10\%. Setting the gravitational
constant equal to $1.1G$ we run the structures for one dynamical
time. This will cause the structures to contract, since a higher value
of $G$ results in a deeper potential. We then set the gravitational
constant equal to $0.9G$ and run the structures for another dynamical
time. This will cause the structures to expand since 
a lower value of $G$ results in a more shallow
potential. All in all we submit each structure to a series of 20
perturbations, alternating between a larger and a smaller value of the
gravitational constant as described. We check that no unwanted oscillatory
behaviour results. We then proceed with ten
perturbations where the value of $G$ is changed by 5\%, and then ten
perturbations where the value of $G$ is changed by only 1\%. 
Finally, we let the structures run for an additional 10
dynamical times with a normal value of $G$ to make sure they are in
equilibrium.

\begin{figure}[h!]
  \resizebox{\hsize}{!}{\includegraphics{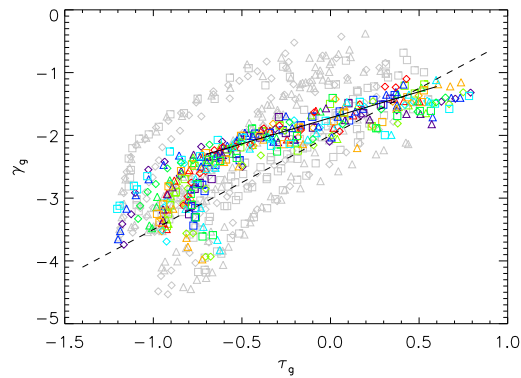}}
  \caption{The same phase space as in Figure \ref{fig1} is now
    populated by the final equilibrated profiles of the perturbed
    structures. Note that the initial conditions (light grey symbols)
    have approached the 1-dimensional line, the attractor, from both
    sides. Also shown is a line of the form $\rho_g/T_g^{3/2} \sim r^{-2}$ 
    (dashed), and a linear guide-the-eye line of the form $\gamma_g = 0.83\tau_g - 1.72$
    in the region $-0.7 < \tau_g < 0.6$ (solid).    
    }
  \label{fig2}
\end{figure}

\section{The Attractor Solution}
The results of the perturbations are presented in Figure \ref{fig2},
from which it is clear that all the structures end up along the same
curve. It is important to emphasise that we had initial conditions
both above and below this attractor solution. The particles in the
outer region (small $\tau$) still tend to follow the initial
conditions. This is partly because they have not been perturbed as
much due to their larger dynamical time.

For the final structures the formed stars dominate the total mass 
within $r = 20$kpc, and they contribute significantly to the collisionless 
component within $r = 100$kpc. To determine how much of an effect 
the star formation has had on 
the attractor, we first perform a simulation in which the density threshold
for star formation, $\rho_{th}$, is ten times larger than normal. This has
the effect of suppressing the star formation. However, the structure still
ends up along the attractor suggesting that the star formation itself does
not have much of an effect.

We also perform a run of only 20 perturbations of 10\%. After that we let the 
structure relax for 30 dynamical times. Again, the structure does not stray
from the attractor at all, showing that it is not affected by the formation of 
stars even over such a long period of time.

To quantify how much the attractor solution stands out compared to the
initial conditions we introduce the structure-to-structure dispersion
$\sigma_{S2S}^2(\tau_g)$, which describes the spread among a group of
structures as a function of $\tau_g$. This value is expected to be
significantly larger for the initial conditions than for the perturbed
structures, because the different structures make clearly distinct lines in Figure \ref{fig1}. 
One can define
\begin{equation}
\sigma_{tot}^2(\tau_g) = \sigma_{S2S}^2(\tau_g) + \sigma_{int}^2(\tau_g),
\end{equation}
where $\sigma_{tot}^2(\tau_g)$ is the total spread of data points, and
$\sigma_{int}^2(\tau_g)$ is the average internal spread of the
individual structures. Perfectly smooth structures with infinitely
many particles would thus have $\sigma_{int}^2(\tau_g) \approx 0$.
Binning the data points with respect to $\tau_g$ we consider
a single bin $i$ with $N$ data points. Since the data points are not distributed symmetrically
around an average value of $\gamma_g$, we define the {median} value of $\gamma_g$ as the
one for which 50\% of the data points lie to either side. Similarly, counting 15.8\% from
the left (one standard deviation) we find {
\begin{equation}
\sigma_-(\tau_g^i) = | \gamma_g^i(50-15.8 \textrm{ percentile}) - \gamma_g^i(50 \textrm{ percentile}) |,
\end{equation}
and counting 15.8\% from the right we find 
\begin{equation}
\sigma_+(\tau_g^i) = | \gamma_g^i(50+15.8 \textrm{ percentile}) - \gamma_g^i(50 \textrm{ percentile}) |.
\end{equation}}
We now define 
\begin{equation}
\sigma_{tot}^2(\tau_g^i) = \left( \frac{\sigma_-(\tau_g^i) + \sigma_+(\tau_g^i)}{2}\right)^2
\end{equation}
and its error
\begin{eqnarray}
\delta \sigma_{tot}^2(\tau_g^i) &=& \frac{1}{2} \left( (\sigma_-(\tau_g^i) - \sigma_{tot}(\tau_g^i))^2 + (\sigma_+(\tau_g^i) - \sigma_{tot}(\tau_g^i))^2\right) \nonumber \\
&=& { \left( \frac{\sigma_+(\tau_g^i) - \sigma_(\tau_g^i)-}{2}\right)^2.}
\end{eqnarray}
For a single structure with $M$ radial bins we define
\begin{equation}
\Delta \gamma_g^j = \gamma_g^j - \hat{\gamma}_g^j
\end{equation}
as the distance between a point $\gamma_g^j$ and its interpolated value
\begin{equation}
\hat{\gamma}_g^j =  \gamma_g^{j-1} + \frac{\gamma_g^{j+1} - \gamma_g^{j-1}}{\tau_g^{j+1} - \tau_g^{j-1}} (\tau_g^j - \tau_g^{j-1}).
\end{equation}
{Here $j$ refers to a given radial bin.} Again, binning in $\tau_g$ we define for a single bin $i$ with $K$ data points the value
of $\sigma_{int}^2(\tau_g^i)$ as the {median} value of $(\Delta \gamma_g)^2$. For the
distribution we again count 15.8\% from left and right and take the average to
to find the error $\delta \sigma_{int}^2(\tau_g^i)$. We do not include any points for which $\gamma_j < -3.5$.

\begin{figure}[h!]
  \resizebox{\hsize}{!}{\includegraphics{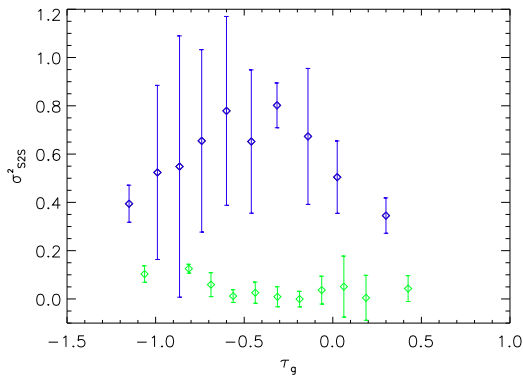}}
  \caption{The structure-to-structure dispersion, describing the
    off-set between a group of different structures.  The initial
    conditions (blue) cover a large fraction of the phase space
    $(\tau_g, \gamma_g)$, whereas the final (green) structures end
    essentially on top of each other, resulting in a very small
    $\sigma_{S2S}^2$. }
  \label{fig3}
\end{figure}

The spread between the initial and final structures are presented in
Figure \ref{fig3}. 
Since the initial conditions span a large portion of the
$\gamma_g-\tau_g$ space they consequently have a relatively large
$\sigma_{S2S}^2$, while the final structures, which generally lie
along the same curve, have a distinctly smaller $\sigma_{S2S}^2$.

\section{Discussion}
In Figure \ref{fig2} we plot a line of the form
$\rho_g/T_g^{3/2} \sim r^{-\alpha}$, corresponding to the entropy
following a power-law in radius, where $\alpha$ is an unknown
constant.  This implies $\gamma_g = -\alpha + 3/2 \, \tau_g$.  
This possibility has been
discussed intensively in the literature 
\citep[see e.g.][for references]{2007MNRAS.380.1521M,
  2010A&A...511A..85P, 2011MNRAS.417.1853D}.  In the figure we use
$\alpha=2$.  Our resulting attractor clearly differs from this simple
behaviour, both in the inner and the outer region.

We also include a linear guide-the-eye line of the form $\gamma_g = 0.83\tau_g - 1.72$.
Although this is obvious a simplified approximation it provides an 
acceptable fit in the range $ -0.7 < \tau_g < 0.6$.

We note that our limited resolution does not allow us to resolve
the inner region where $\gamma_g > -1$. That region is particularly
interesting when one is considering cool-core clusters
where $\tau_g$ may be large and positive. Initially the rapid cooling 
in our simulations forms a large central stellar condensation, which acts like a 
(unrealistically large) central blue BCG, a problem which possibly could be 
reduced by including AGN feedback. Subsequently, our perturbations are 
likely inducing adiabatic compression shocks, which helps (together with the 
SN feedback) preventing the structures from undergoing further rapid cooling, 
in fact, we do not observe any late time cooling catastrophe despite our simulation 
effectively being longer than a Hubble time. This may also be connected 
to the fact that the quickly formed central BCG may act like a stabilizing
structure which helps preventing rapid over cooling from destabilizing the
structure \citep{li2011}.

This simple connection allows one to write a simplified hydrostatic 
equilibrium equation,
$M_{\rm tot} = f_1(T_g)$,
where $f_1(T_g)$ is a known function only of the gas temperature, with
no reference to the gas density. Alternatively, if we knew the absolute
value of the temperature at one point, then one could write this as
$M_{\rm tot} = f_2(\rho_g)$, where $f_2(\rho_g)$ is a known function only of the gas 
density, or even better, 
$M_{\rm tot} = f_3(I)$, where $f_3(I)$ is a known function only of the gas 
intensity, which is easily observable even at high redshift. 

It has long been observed that the characteristic radius for
the temperature (at $T_{\rm max}$) roughly corresponds to
the scale radius for the gas density~\citep[see e.g.][]{2006ApJ...640..691V}. 
We now see how this relates
to the attractor presented here: The radius corresponding to
$\tau_g = 0$ should have $-2 < \gamma_g< -1.5$, which is
close to the scale radius of the density profile 

A connection has previously been suggested between the gas temperature
profile and the dark matter dispersion~\citep{2007A&A...476L..37H}, 
which is not only a fair approximation for galaxy 
clusters~\citep{2009ApJ...700.1603F,2009ApJ...690..358H,2011ApJ...728...54Z},
but also holds well for the hot gas in
galaxies as well as near AGN~\citep{2011ApJ...734...62H}.
We find that the corresponding structure-to-structure dispersion
for this gas-DM temperature connection is larger, 
and we therefore conclude that the gas-density
connection presented in this paper is likely more fundamental.

When comparing our finding here with the recent discovery of an attractor 
for dark matter structures, a  remarkable suggestion presents itself. The
dark matter attractor is a 1-dimensional line in a 3-dimensional space
spanned by $\gamma_{dm}, \tau_{dm}$ and $\beta_{dm}$, where the first two
are the logarithmic derivatives of the dark matter density and radial velocity
dispersion ($\sigma_r^2$, which acts like a temperature per mass), and $\beta_{dm}$ is the 
velocity anisotropy~\citep{2010ApJ...718L..68H}.
The Jeans equation, which is the dark matter equivalent of the
hydrostatic equilibrium, can be written as
\begin{equation}
  M_{tot}(r) = -r \frac{\sigma_r^2}{G} \left(\gamma_{dm}(r) + \tau_{dm}(r) 
+2 \beta_{dm}(r)\right) \, .
\label{eqn:jeans}
\end{equation}
We therefore see that the dark matter attractor is between elements in
the parenthesis in the Jeans equation, eq.~(\ref{eqn:jeans}), and the
gas attractor presented in this paper is between the elements in the
parenthesis in the equation for hydrostatic equilibrium,
eq.~(\ref{eqn:he}). We therefore speculate, that if one will manage to
derive the gas attractor we present here, then one may possibly use that
understanding to  derive the attractor for the dark matter as well.

It is important to emphasize that the attractor for the gas is not just
an reaction to the changes in the dark matter as it adjusts to its own 
attractor. Two of our initial structures start right on the dark matter
attractor but not on the gas attractor. When they are perturbed they
move to the gas attractor while staying on the dark matter attractor.

In this paper we have only investigated the effect of the particular
$G$-perturbation. We note that the resulting dark matter density profiles
are different, and the gas attractor is thus not a trivial result
of the perturbation driving the dark matter structures towards a universal
density profile.
It has been shown for the dark matter attractor that a
wide range of perturbations all lead to the same attractor 
\citep{2010ApJ...718L..68H, sparre12, barber12}, suggesting
that our found gas attractor is likely not a result of the particular
perturbation chosen here.
In the future we intend to study which effects
more realistic perturbations, such as minor or major mergers, will
have on the gas attractor.
We also note that the simulations presented here both have gas and
dark matter, however, when analysing the 3-dimensional dark matter phase
space, we find that the dark matter attractor is slightly shifted
compared to the cases including only dark matter.

\section{Conclusions}
Using high-resolution numerical simulations of large galaxy clusters,
including radiative cooling, star-formation and feedback, we identify
an attractor for the hot gas component.  This attractor gives a very
simple connection between the temperature and the density of the gas.

We find the attractor to hold for all structures which have been
exposed to significant perturbations and subsequently allowed to
relax. Since cosmological structures experience 
perturbations during mergers, we speculate that the central part of
galaxy clusters, which are fully equilibrated, may follow the attractor. It
will be very interesting in the near future to test this suggestion on
real X-ray observations.

\acknowledgments
We are pleased to thank Teddy Frederiksen, Ole H{\o}st, 
Marco Roncadelli, and Martin Sparre
for useful discussions. The simulations were performed on the facilities provided
by the Danish Center for Scientific Computing.
The Dark Cosmology Centre is funded by the Danish National Research Foundation.

\clearpage




\clearpage


\begin{thebibliography}{}


\bibitem[Allen et 
al.(2011)]{2011ARA&A..49..409A} Allen, S.~W., Evrard, A.~E., \& Mantz, A.~B.\ 2011, \araa, 49, 409 

\bibitem[Barber et al.(2012)]{barber12} Barber, J.\ et al.\ 2012, to appear

\bibitem[Binney 
\& Tremaine(2008)]{2008gady.book.....B} Binney, J., \& Tremaine, S.\ 2008, Galactic Dynamics: Second Edition.
Princeton University Press

\bibitem[Buote 
\& Lewis(2004)]{2004ApJ...604..116B} Buote, D.~A., \& Lewis, A.~D.\ 2004, \apj, 604, 116 

\bibitem[Cavaliere 
\& Fusco-Femiano(1976)]{1976A&A....49..137C} Cavaliere, A., \& Fusco-Femiano, R.\ 1976, \aap, 49, 137 

\bibitem[Dubois et al.(2011)]{2011MNRAS.417.1853D} Dubois, Y., Devriendt, 
J., Teyssier, R., \& Slyz, A.\ 2011, \mnras, 417, 1853 
\bibitem[Frederiksen et al.(2009)]{2009ApJ...700.1603F} Frederiksen, T.~F., 
Hansen, S.~H., Host, O., \& Roncadelli, M.\ 2009, \apj, 700, 1603 

\bibitem[Hansen 
\& Piffaretti(2007)]{2007A&A...476L..37H} Hansen, S.~H., \& Piffaretti, R.\ 2007, \aap, 476, L37

\bibitem[Hansen, Juncher \& Sparre(2010)]{2010ApJ...718L..68H} Hansen, S.~H., Juncher, 
D., \& Sparre, M.\ 2010, \apjl, 718, L68

\bibitem[Hansen et al.(2011)]{2011ApJ...734...62H} Hansen, S.~H., 
Macci{\`o}, A.~V., Romano-Diaz, E., et al.\ 2011, \apj, 734, 62 

\bibitem[Host et al.(2009)]{2009ApJ...690..358H} Host, O., Hansen, S.~H.,
Piffaretti, R., et al.\ 2009, \apj, 690, 358

\bibitem[Host 
\& Hansen(2011)]{2011ApJ...736...52H} Host, O., \& Hansen, S.~H.\ 2011, \apj, 736, 52 

\bibitem[Katz et al.(1996)]{1996ApJS..105...19K} Katz, N., Weinberg, D.~H., 
\& Hernquist, L.\ 1996, \apjs, 105, 19 

\bibitem[Kennicutt(1998)]{1998ApJ...498..541K} Kennicutt, R.~C., Jr.\ 1998, 
\apj, 498, 541

\bibitem[Kneib et al.(2004)]{2004ApJ...607..697K} Kneib, J.-P., Ellis, 
R.~S., Santos, M.~R., \& Richard, J.\ 2004, \apj, 607, 697 

\bibitem[Li \& Bryan(2011)]{li2011}Li, Y., \& Bryan, G. L. \ 2011

\bibitem[Lynden-Bell(1967)]{1967MNRAS.136..101L} Lynden-Bell, D.\ 1967, 
\mnras, 136, 101 

\bibitem[Morandi 
\& Ettori(2007)]{2007MNRAS.380.1521M} Morandi, A., \& Ettori, S.\ 2007, \mnras, 380, 1521 

\bibitem[Moster et al.(2011)]{2011MNRAS.415.3750M} Moster, B.~P., 
Macci{\`o}, A.~V., Somerville, R.~S., Naab, T., 
\& Cox, T.~J.\ 2011, \mnras, 415, 3750 

\bibitem[Pointecouteau et 
al.(2005)]{2005A&A...435....1P} Pointecouteau, E., Arnaud, M., \& Pratt, G.~W.\ 2005, \aap, 435, 1 

\bibitem[Pratt et 
al.(2010)]{2010A&A...511A..85P} Pratt, G.~W., Arnaud, M., Piffaretti, R., et al.\ 2010, \aap, 511, A85 

\bibitem[Sarazin(1986)]{1986RvMP...58....1S} Sarazin, C.~L.\ 1986, Reviews 
of Modern Physics, 58, 1 

\bibitem[Sparre \& Hansen(2012)]{sparre12} Sparre, M., \& Hansen, S.~H.\ 2012, to appear

\bibitem[Springel 
\& Hernquist(2003)]{2003MNRAS.339..289S} Springel, V., \& Hernquist, L.\ 2003, \mnras, 339, 289 

\bibitem[Springel(2005)]{2005MNRAS.364.1105S} Springel, V.\ 2005, \mnras, 
364, 1105 

\bibitem[Vikhlinin et al.(2006)]{2006ApJ...640..691V} Vikhlinin, A., 
Kravtsov, A., Forman, W., et al.\ 2006, \apj, 640, 691 

\bibitem[Vikhlinin et al.(2009)]{2009ApJ...692.1060V} Vikhlinin, A., 
Kravtsov, A.~V., Burenin, R.~A., et al.\ 2009, \apj, 692, 1060 

\bibitem[ZuHone(2011)]{2011ApJ...728...54Z} ZuHone, J.~A.\ 2011, \apj, 728, 
54 




\end{thebibliography}
\end{document}